\documentclass[a4paper]{fcs}
\usepackage{bm}
\usepackage{mathrsfs}
\usepackage{threeparttable}
\usepackage{multirow}
\usepackage{graphicx}
\usepackage{amsmath}
\usepackage{caption}
\usepackage{subcaption}

\graphicspath{{images/}}
\volumn{ }
\doi{ }
\articletype{RESEARCH~ARTICLE}
\copynote{{\copyright}add copy right}
\ratime{Received month 06, 2018}
\email{bsse0622@iit.du.ac.bd}

\title{A Regression-Based Share Market Prediction Model for Bangladesh}
\author{Syeda Tasnim Fabiha \xff, Rubaiyat Jahan Mumu \xff,  Farzana Aktar\xff \\ B M Mainul Hossain}
\address{{Institute of Information Technology, University of Dhaka, Dhaka, Bangladesh}}

\begin{document}
\maketitle
\setcounter{page}{1}
\setlength{\baselineskip}{14pt}

\begin{abstract}
Share market is one of the most important sectors of economic development of a country. Everyday almost all companies issue their shares and investors buy and sell shares of these companies. Generally investors want to buy shares of the companies whose market liquidity is comparatively greater. Market liquidity depends on the average price of a share. In this paper, a thorough linear regression analysis has been performed on the stock market data of Dhaka Stock Exchange. Later, the linear model has been compared with random forest based on different metrics showing better results for random forest model. However, the amount of individual significance of different factors on the variability of stock price has been identified and explained. This paper also shows that the time series data is not capable of generating a predictive linear model for analysis.
           
\end{abstract}


\section{Introduction}
\label{intro}
\noindent Development of a nation's capital market often reflects its economic development. As a component of this market, the stock market, which is also known as share market, plays an important role in industrial development through supply of capital. The importance of institutions like the stock exchanges cannot be overemphasized for the development of capital market. Share market is a place where shares of different companies are allowed to be brought and sold. Bonds, mutual funds and derivative contracts are also traded along with share of the companies. There are two types of share markets. One type is called the primary share market and the other type is secondary share market. In primary share market, a company get registered to issue shares to the public and thus raise money. Generally, companies are get listed on the stock exchange through primary market route. In secondary share market, investors buy shares from other investors at prevailing price.

The capital market is vital to the long-run growth and prosperity of the business sector, as it facilitates the transfer of funds from savers to investors. By requiring the disclosure of certain corporate financial data, an efficient market system allows investors to assess the risk-return trade-offs involved in a transaction and move funds towards comparatively more promising investments. The secondary markets allows investors and financial institutions to alter the liquidity, composition and risks of their portfolios in response to newer information on changes in market condition. A number of brokers conduct these transactions in secondary share market. An investor may sell all its shares and exit the financial market if it wants. Every investor wants to buy share at lowest price and sell at comparatively higher price. Therefore, investors want to predict the market change rate earlier. Those, who can predict the market, are the gainers. 

In Bangladesh, investors become more interested in the secondary market for its liquidity, and the inherent return possibility due to the volatility of the market. The first stock market operator of Bangladesh is "Dhaka Stock Exchange", was set up in 1954. To integrate the entire economic role, it was reactivated in 1976 with the listing of nine companies. The capital market of Bangladesh has already experienced some major collapse 2010 \cite{collapse2010}. The incident was popularly known as "2010-11 Bangladesh Share Market Scam". The major reasons of this incident was entrance of inexperienced investors and the sudden unsustainable growth of stock market in a very short period. This research has focused on the features and factors of a company which affects company movement. For this analysis, weekly time series data of certain companies has been collected from January, 2012 to May, 2015. A regression analysis has been conducted on these data and a report is generated finally. 

A large number of studies had been conducted on the status and determinates of stock exchange market. A study was conducted focusing towards identifying the factors of stock price volatility and these effects on A-rated companies and B-rated companies \cite{ahmed2014factors}. It used multiple linear regression model using SPSS to determine the statistical significance of the independent variables. Another study was conducted to see the effectiveness of 17 variables for the movement of stock price in Bangladesh capital market by using SPSS. It was also conducted on Dhaka Stock Exchange of Bangladesh (DSEBd). Later this study found 5 core factors affecting price movement. \cite{islam2015factors}.

As accurate stock market prediction is one of the major problem for investors, a study used neural networks to produce a successful model that could be used for stock market prediction. The prediction system was made up of several neural networks that learned the relationships between various technical and economical indexes and the timing for when to buy and sell stocks. The goal of that study was to predict the best time to buy and sell for one month in the future \cite{kimoto1990stock}. Researches on Karachi Stock Exchange (KSE) was done to predict the market performance on day closing using different machine learning techniques. It compared several machine learning techniques including Single Layer Perceptron (SLP),  Radial Basis Function (RBF) and Support Vector Machine (SVM) \cite{usmani2016stock}.

The previous works on stock market prediction considered various multi-dimensional factors for generating a predictive model. However, there have not been many works on the thorough analysis of stock market data of Bangladesh. In this study we propose to run a predictive analysis for stock market price of Dhaka Stock Exchange. As a result of analysis, we will be able to tell the amount of effect of each factor on the variability of the stock price and predict future stock prices of the companies with a significant confidence level. 

\section{Methodology}
\label{method}
\noindent For generating an efficient predictive model for the stock market dataset, two different types of methods are being followed in this work. One is parametric: linear regression and another is non-parametric: random forest. Initially all the predictors have been considered as important contributors for the model generation. Later, analysis is performed on the identification of the best predictors, which contribute most for describing the variability of the stock closing prices. 

\subsection{Linear Regression}
Linear regression is an approach of linear modeling between a dependent variable Y and one or more independent variables. If the number of independent variable is one, then it is called simple linear regression. The example equation for simple linear regression is given below:
\begin{equation} \label{simple linear equation}
    \hat{y}=\beta_{0} + \beta_{1} x
\end{equation}

where $\hat{y}$ is the response variable, $x$ is the predictor, $c$ is the intercept and $\beta$ is the slope of a linear equation. 

When linear Regression is performed for multiple predictors, it is Multiple Linear Regression. It gives each predictor a separate slope coefficient. If we have n distinct predictors. Then multiple linear regression model is:

\begin{equation}
\hat{y}=\beta_{0} x_{0}+\beta_{1} x_{1}+\beta_{2} x_{2}+ \cdots +\beta_{n} x_{n}+ \epsilon
\end{equation}

Here, $\epsilon$ is the error term in regression models. In a linear Regression model, attempt is made to predict $\beta_{0}$, $\beta_{1}$, $\cdots$, $\beta_{n}$ by minimizing the Residual Sum of Squares (RSS) using the Least-Squares method. 

\begin{equation} \label{rss1}
RSS= \sum_{i=1}^{k} (y_{i}-\hat{y_{i}})^{2}
\end{equation}

Like the simple linear regression, the multiple linear regression also aims to reduce RSS by substituting the value of $\hat{y}$ into Equation \ref{rss1}.

\begin{equation} \label{rss2}
\resizebox{0.9\columnwidth}{!}{$RSS = \sum_{i=1}^{k} (y_{i}-(\beta_{1} x_{1}+\\beta_{2} x_{2}+\beta_{3} x_{3}+ \cdots +\beta_{n} x_{n}+c))$}
\end{equation}

After fitting the linear regression model to a particular data-set, some problems may arise based on different data properties such as non-linearity of response-predictor relationships, non-constant variance of error terms, out-liars, high leverage points, co-linearity etc. These problems can result in small R square and t values and bigger p values indicating lack of significant relationship between the predictors and response. To overcome these problems, different data manipulation techniques are performed on our data-set such as transformation ($\log{ x}$, $\sqrt{x}$, ${x}^2$), use of weighted least squares, removing out-liars, high leverage points, dropping or combining co-linear variables using Variance Inflation Factor (VIF) etc. 

\subsection{Random Forest}
This tree-based approach involves producing multiple trees that are combined to generate a single prediction with probable high accuracy. In order to avoid the high variance problem of decision trees, bagging increases the prediction accuracy at the expense of loss of interpretability. Random Forest is an improvement for the bagging concept \cite{breiman2001random}. Instead of bootstrapping of training samples used by Bagging, a random sample of m predictors is chosen as split candidates from the full set of n predictors where typically $ m \approx \sqrt{n}$. 

\subsection{Evaluation Metrics}

Three metrics Mean Absolute Error (MAE), Root Mean Square Error (RMSE), Pearson Correlation Coefficient are used to compare the two above-mentioned approaches for model generation. 

\subsubsection{MAE}

MAE gives the average magnitude of errors in a set of predictions without considering their direction.  It’s the average over the test sample of the absolute differences between prediction and actual observation where all individual differences have equal weight.

\begin{equation} \label{mae}
MAE = \frac{1}{n}\sum_{j=1}^{n} (y_{j}-\hat{y_{j}})
\end{equation}

\subsubsection{RMSE}

This metric also calculates average magnitude of error. The errors are squared before they are averaged giving the large errors relatively higher weights. This means RMSE should be more useful when large errors are particularly undesirable.

\begin{equation} \label{rmse}
RMSE = \sqrt{\frac{1}{n}\sum_{j=1}^{n} (y_{j}-\hat{y_{j}})^{2}}
\end{equation}

\subsubsection{Pearson Correlation Coefficient}

This metric gives an idea about the strength of relationship between the response and the predictors. The amount of variability in the responses is explained using correlation coefficient. When there are multiple predictors, then it is termed as R square.

\vspace{1cm}

\subsection{Support Vector Machine}

\section{Experimentation}
\noindent In this section the data description is provided after being explored. Moreover, the Paired T-Test performed to select the best predictor is also presented.
\subsection{Data}
\label{datas}
Most of the data was collected from the Stock market related websites of Bangladesh like Lanka Bangla Financial Portal, Dhaka Stock Exchange Bangladesh Ltd. and Trading Economics \cite{lankabd} \cite{dsebd} \cite{trade}. From these sites, stock market related data from 2013 to 2017 for five pharmaceutical companies are collected and aggregated to form the training data. Data was collected on quarter basis from 4th quarter of 2013 to 1st quarter of 2017.  The five selected companies are: ACI Pharmaceuticals, Beximco Pharmaceuticals Ltd, Renata Limited, Glaxo SmithKline Bangladesh Ltd and Square Pharmaceuticals Limited. The descriptions of the predictors and the response are given in Table \ref{data}.
\begin{table*}[!ht]
\centering
\caption{Predictor Name and Data Description}
\label{data}

\begin{tabular}{|p{1.2cm}|p{2.5cm}|p{2cm}|p{8.5cm}|}
\hline
\textbf{Type} & \textbf{Predictor Name} & \textbf{Acronym} & \textbf{Data Description}   \\ \hline
\multirow{9}{*}{Predictor} & Term & Term & Term means the quarter number  \\ \cline{2-4} 
                           & Stock Price & Stock Price  & Stock price means the price of each share of a company  \\ \cline{2-4} 
                           & Return on Asset & ROA & Return on assets (ROA) is an indicates the profitability of a company related to its total assets. \\ \cline{2-4} 
                           & Return on Equity & ROE & Return on equity indicates the net income returned as a percentage of shareholders equity. \\ \cline{2-4} 
                           & Current Ratio & CR & The current ratio means the liquidity ratio to measure a company's ability to pay short-term and long-term obligations \\ \cline{2-4} 
                           & Total Asset Turn Over & TATO & Asset turnover is the the value of a company's sales or revenues generated relative to the value of its assets \\ \cline{2-4} 
                           & Debt to Total Asset & DTA & Total debt to total assets is a leverage ratio that defines the total amount of debt relative to assets \\ \cline{2-4} 
                            & Panel & Panel & Panel is the unique code for each company  \\ \hline

\end{tabular}
\end{table*}
Stock prices change because of supply and demand. If more people want to buy a stock than sell it, then the price scales up. Usually, best performing  companies tend to have increased stock price. If a company is earning is better than expected, the stock price jumps up. The possible predictors are shown in table \ref{data}. This is a time series dataset i.e. readings of the same predictors are taken at unit intervals of time. The current ratio (CR) considers the current total assets of a company (both liquid and non-liquid) relative to that company's current total liabilities. The current ratio is also known as the working capital ratio. The Asset Turnover ratio is used as an indicator of the efficiency of a company in deploying its assets for revenue generation.


\begin{table}[!t]
\centering
\caption{Coefficients of Linear Regression Analysis (****p<0.001, ***p<0.01, **p<0.05, *p<0.1)}
\label{sigtab}
\resizebox{\columnwidth}{!}{ 
\begin{tabular}{|c|c|c|c|}
\hline
\textbf{Predictors} & \textbf{F-stat} & \textbf{$R^2$} & \textbf{RSE}  \\ \hline
Panel         & 1.861           & 0.02783            & 565.20                             \\ \hline
DTA                  & 33.85         & 0.3425            & 464.80                                \\ \hline
ROE                  & 14.59         & 0.1834             & 518.00                                \\ \hline
ROA                 & 2.122          & 0.03027            & 594.00                               \\ \hline
TATO               & 67.9         & 0.5109             & 400                               \\ \hline
CR                 & 5.272         & 0.7506            & 551.30                               \\ \hline
Term                  & 0.7566          & 0.01151            & 569.00                               \\ \hline

\end{tabular}
}
\vspace{-0.55cm}
\end{table}

\begin{figure*}[!ht]
\centering
\includegraphics[width=15cm,height=10cm]{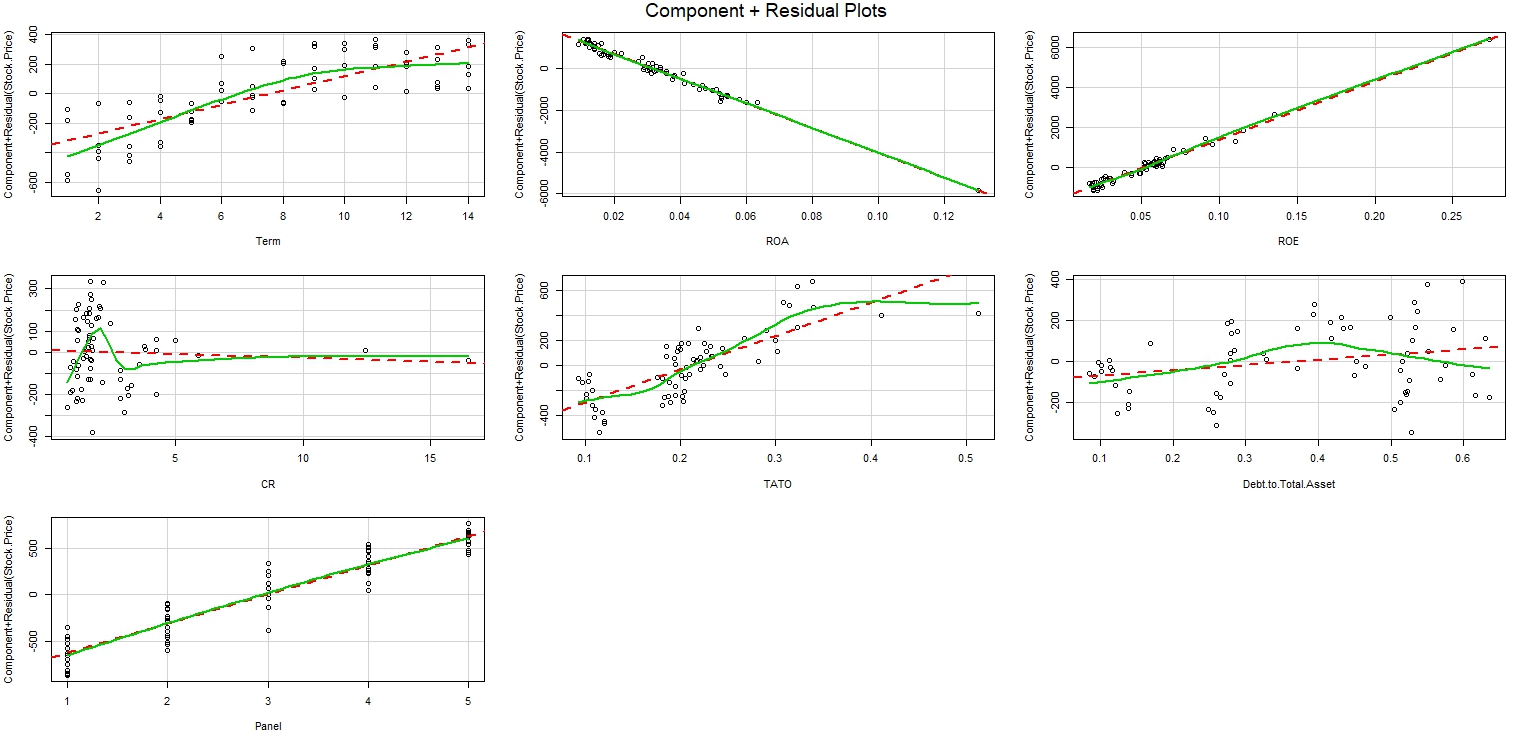}
\vspace{-0.3cm}
\caption{Component Residual Plots for the Predictors}
\label{cr}
\vspace*{-0.3cm}
\end{figure*}

Table \ref{sigtab} shows F-statistic, R square and residual sum of squares, RSE values. From the data, TATO, DTA, ROE have comparatively larger F-statistic value. This indicates the existence of relationship between the response and the predictors. RSE estimates the standard deviation of the response from the population regression line. These values indicate the amount of error with respect to the mean values.R square statistic records the percentage of variability in the response that is explained by the predictors. \begin{figure*}[!htbp]

\centering
\begin{subfigure}[t]{0.5\textwidth}
\centering
\includegraphics[width=\linewidth,keepaspectratio]{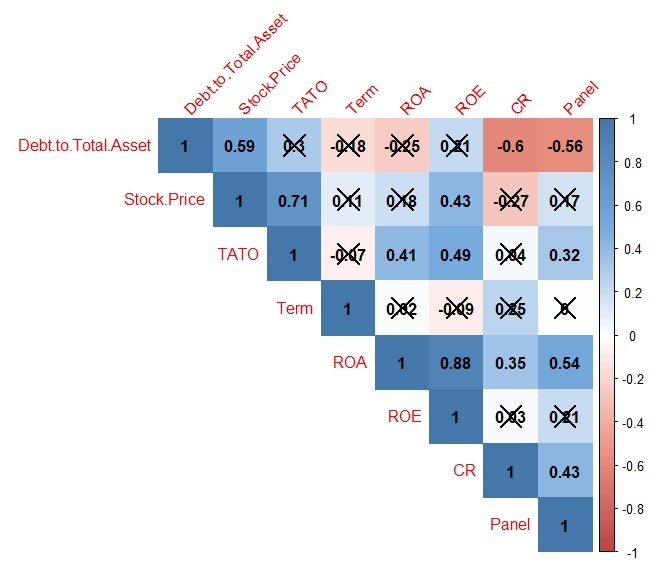}
\caption{0.05 Significance Level}
\label{cor1}
\end{subfigure}%
\begin{subfigure}[t]{0.5\textwidth}
\centering
\includegraphics[width=\linewidth,keepaspectratio]{cor2.jpeg}
\caption{0.01 Significance Level}
\label{cor2}
\end{subfigure}

\caption{Pearson correlation of the Features (Non-significant features are marked with a cross mark (X))}
\label{cor}
\vspace*{-0.3cm}
\end{figure*}

Figure \ref{cor} shows correlation among the predictors of the dataset. It is evident that three predictors DTA, TOTA, ROE are significant relationship with stock price each having correlation value 0.59, .71 and .43 respectively.

\section{Fitting to Linear Model}
\noindent Figure \ref{cr} shows the component residual plots for each of the predictors. It is evident that in most cases, the difference between residual line and component line is small which indicates linearity.However, for Term, CR, TATO, DTA the difference is larger. Therefore, these predictors are transformed
for a linear relationship according to Tukey power transformation \cite{tukey}. As a solution to this issue, we performed transformation and found that neither using square root nor taking log value improved results. Therefore, no transformation was applied on the predictors.

\begin{figure}[!th]
\centering
    \includegraphics[width=\columnwidth,height=7cm]{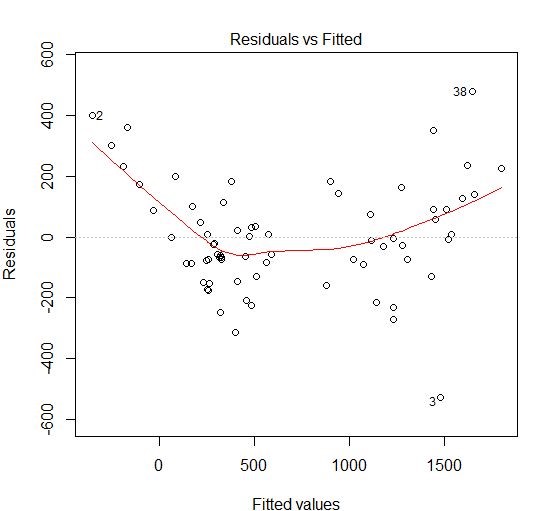}
    \vspace{-0.7cm}
    \caption{Regression Diagnostic Plots}
    \label{residual}
\end{figure}

Observing the residual versus fitted plot shown in figure \ref{residual}, we find no significant pattern. This shows evidence of linearity in the model. Also after removal of outliers, the plot generated more effective results.From the plot, we do not observe any funnel shape that represents absence of non-constant variance of errors in the data called Heteroscedasticity. Therefore, the model satisfies the homoscedasticity property of fitted Linear Model.After log-transforming the data, we found the shape disappeared leaving some outliers. Three outliers 3, 38, 48 were found and removed from the dataset. 

\begin{figure*}[!htbp]
\centering
\begin{subfigure}[t]{0.5\textwidth}
\centering
\includegraphics[width=\linewidth,keepaspectratio]{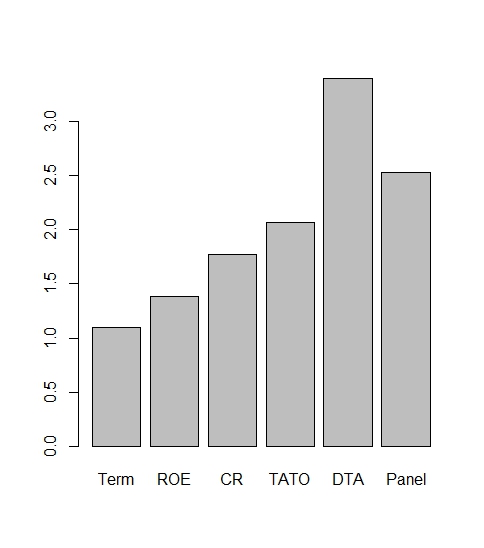}
\caption{Variance Inflation Factors of the Predictors}
\label{vif}
\end{subfigure}%
\begin{subfigure}[t]{0.5\textwidth}
\centering
\includegraphics[width=\linewidth,height = 10cm]{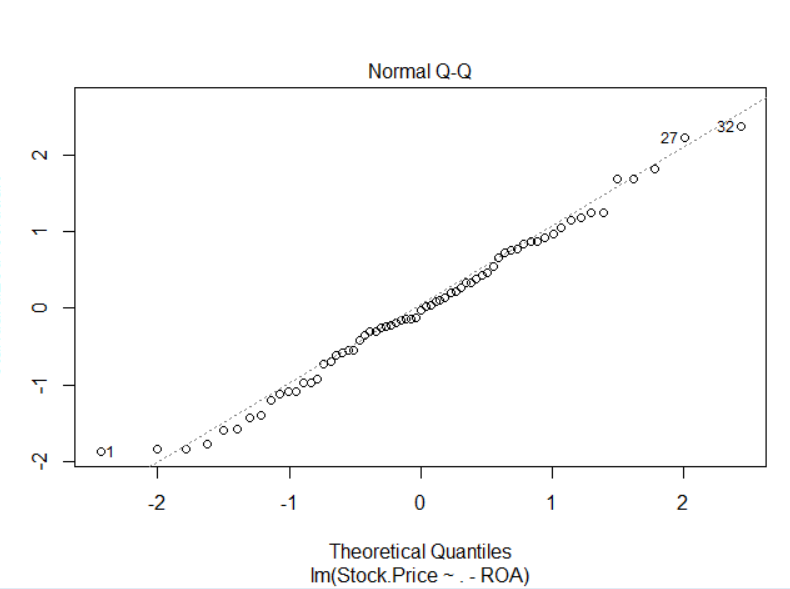}
\caption{Regression Diagnostic Plots}
\label{cor2}
\end{subfigure}
\caption{VIF and Normal Q-Q}
\label{qq}
\vspace*{-0.3cm}
\end{figure*}

In figure \ref{vif} the Variance Inflation Factor (VIF) values for the predictors including ROA generated very high values for ROA and ROE. This means that the ROA and ROE values must be collinear. We excluded ROA from the predictor set. Then the VIF values of all the predictors became less than 4 indicating absence of multicollinearity in the model. From figure \ref{qq}b, normal Q-Q plot shows that standardized residuals closely follow a linear trend. This confirms normality of residuals. 

\begin{figure}[!ht]
\centering
    \includegraphics[width=\columnwidth,height=7cm]{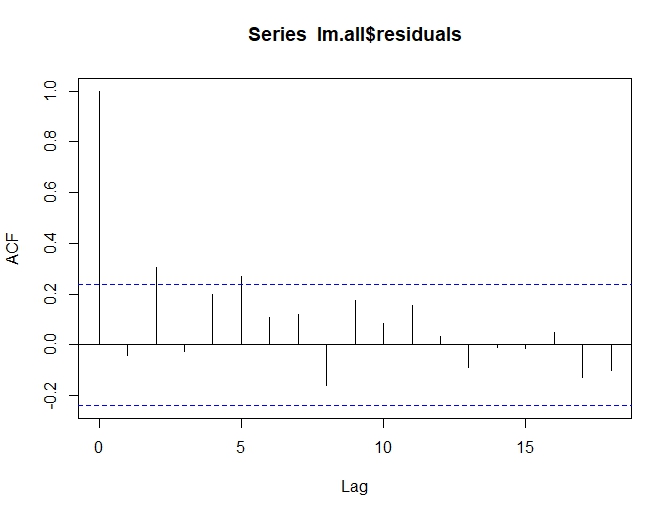}
    \vspace{-0.7cm}
    \caption{Autocovariance and Autocorrelation Functions Plot}
    \label{autocor}
\end{figure}

Figure \ref{autocor} shows the autocorrelation function plot for residual time series. The first line indicates the correlation of the residual with itself. This is why it is larger. The next lags of the residuals are small and down the dotted blue line, which indicates that current values of residuals are not dependent on previous values.

After we are done with data pre-processing to fit them into a linear model, it is found that our model satisfies the conditions of linear modelling such as measurement error free observation of predictors and responses, linearity, normality, homoscedasticity, no multicollinearity and no autocorrelation.

\section{Application of Evaluation Metrics}
\begin{table}[t!]
\centering
\caption{MAE, RMSE and Correlation Coefficient values (* indicates statistically significant at 0.05 level compared to Linear Regression)}
\label{exp_tab}
\resizebox{\columnwidth}{!}{%
\begin{tabular}{|p{1.8cm}|p{1.8cm}|p{1.8cm}|p{1.8cm}|}
\hline
\textbf{Metrics}        & \textbf{Linear Regression} & \textbf{Support Vector Machine} & \textbf{Random Forest} \\ \hline
MAE                     & 220.8                       &  -                          & 119.9                   \\ \hline
RMSE                    & 283.7                       &   -                         & 195.9                   \\ \hline
Correlation Coefficient & 0.88                        &   -                        & 0.9                   \\ \hline
\end{tabular}%
}
\vspace{-0.4cm}
\end{table}
\noindent Though the MAE, RMSE and Correlation coefficient values for linear regression and random forest both showed significant results. On this dataset, random forest outperforms linear regression as showed in table \ref{exp_tab}.

\begin{figure*}[!ht]
\centering
\includegraphics[width=15cm,height=10cm]{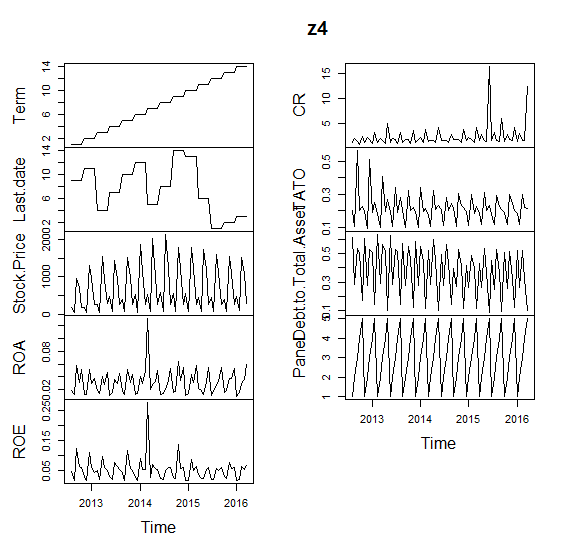}
\vspace{-0.3cm}
\caption{Time Series Plot}
\label{timeseries}
\vspace*{-0.3cm}
\end{figure*}

The possible reason behind this is that the stock market data is a time series data as shown in figure \ref{timeseries}. This kind of time-series data have inherent collinearity among predictors \cite{hamilton} \cite{time_series} \cite{hurvich}. Because of this, the model is more biased towards having better performance using tree based method random forest.

\section{Results and Discussion}
\noindent As discussed previously, the data is fit to a linear regression model. To evaluate the model, 10-fold cross validation is performed. The R2 is 0.88 that indicates data fits well to the linear model.
\begin{table}[!th]
\centering
\caption{Coefficients of Linear Regression Analysis with Principal Components (***p<0.0001, *p<0.05)}
\label{sigtabpca}
\resizebox{\columnwidth}{!}{%
\begin{tabular}{|l|l|l|l|l|}
\hline
\textbf{Predictors} & \textbf{Estimate} & \textbf{Std. Error} & \textbf{t value} & \textbf{Pr(\textgreater|t|)} \\ \hline
(Intercept)         & -1691.838           & 167.744             & -10.086            & 1.55$e^{-14}$***                          \\ \hline
Term                 & 42.113          & 7.348             & 5.731            & 3.44$e^{-07}$ ***                   \\ \hline
ROE                 & 342.091          & 902.390             & 0.379            & 0.706                      \\ \hline
CR                 & -32.597          & 15.769             & -2.067            & 0.403                     \\ \hline
TATO                 & 2367.872           & 824.498             & 8.515             & 3.02$e^{-05}$                      \\ \hline
DTA                 & 2536.810             & 308.046             & 8.235            & 1.95$e^{-11}$***                      \\ \hline
Panel                 & 213.175             & 31.151             & 6.843            & 4.63$e^{-11}$***                      \\ \hline
\end{tabular}%
}
\vspace{-0.6cm}
\end{table}
\vskip 0.30cm
The F-Statistic is 56.21 on 6 and 60 degree of freedom. The p-value of the F-Test is 2.2e-16 which is much less than the 0.05 confidence interval. For this, the null hypothesis that the fit of the intercept-only model and the fitted model is equal, can be rejected. 
Table \ref{sigtabpca} shows the linear regression results. The null hypothesis for the t-test is that there is no relationship between the predictor and the response. For p values 0.001, 0.01, 0.05 and 0.1, we observe if any value is larger than t. It is evident that we can reject the null hypothesis in case of most of the predictors. At 0.001 confidence level, Term, TATO, DTA are Panel are related to the response. This means that, time sequence of stock data, total asset turnover of a company, Debt on total asset and company type impact the stock price of a day. The signs of their slope estimates indicate that stock price changes in both direction with the changes of these variables.

In this regression analysis, ROE estimate is positive showing the increased stock price of shares results in generation of more profit from the money shareholders have invested. However, stock price does not depend much on the amount of profit a company is making because the relationship is reverse. Increased stock price may lead to greater profitability. The estimate of CR is negative also explains the fact clearly that having large values for CR decreases stock price of shares to make greater amount of sales. Generally, the higher the asset turnover ratio, the better the company is performing, the greater is its demand of shares depicting significant increase in its stock price. In this analysis, the maximum estimate is achieved for debt to total asset (DTA) value. The reason behind this can be explained this way, suppose, a company having large debt to total asset ratio tends to be on the list of companies holding greater potential of investment opportunities. This increases their amount of debt though depicting potential lack of financial flexibility. Thus, shareholders tend to buy shares from these companies creating a high demand for their shares and thus increasing the stick price.

From these findings, we get some general decision criteria for predicting stock prices of share market. Due to fast changing rate of stock price, it is very difficult to predict exact stock price. However, the total asset turnover and debt to asset value indicating the current value of a company are potential predictors for predicting the stock price. Again, the main theory of price movement of a stock is what investors feel a company is worth. Therefore, though we can make some idea about public emotion at a given time, it is not possible to quantify this kind of variable.
\vskip 0.30cm

\section{Conclusion}
This paper presents a price predictive model for Bangladesh for predicting stock price and analyzing influence of multiple factors on stock price. It has been shown that considering the most significant factors, prediction can be done with an acceptable accuracy measuring 119.9 in MAE, 195.9 in RMSE and 0.9 in Correlation Coefficient using a random forest model. Regression analysis results show that factors related to return on equity, total asset turnover, debt to total asset,  company sector are the most significant ones. So, the specific factors belonging to these classes can be influenced to predict stock price of companies' shares in Bangladesh. 
\noindent 




\bibliographystyle{unsrt}
\bibliography{main}

\end{document}